\documentclass[aps,prb,superscriptaddress,twocolumn,floatfix,showpacs]{revtex4}
\usepackage{graphicx}
\usepackage[]{amsmath}

\begin{document}

\title{Magnetic phase separation in \chem{EuB_6} detected by muon spin rotation}

\author{M. L. Brooks}
\affiliation{Clarendon Laboratory, University of Oxford, Parks Road, Oxford OX1
3PU, United Kingdom}

\author{T. Lancaster}
\affiliation{Clarendon Laboratory, University of Oxford, Parks Road, Oxford OX1
3PU, United Kingdom}

\author{S. J. Blundell}
\affiliation{Clarendon Laboratory, University of Oxford, Parks Road, Oxford OX1
3PU, United Kingdom}

\author{W. Hayes}
\affiliation{Clarendon Laboratory, University of Oxford, Parks Road, Oxford OX1
3PU, United Kingdom}

\author{F. L. Pratt}
\affiliation{ISIS Muon Facility, ISIS, Chilton, Oxon. OX11 0QX, United Kingdom}

\author{Z. Fisk}
\affiliation{Department of Physics, University of California, Davis, Davis, CA
95616}

\date{\today}

\newcommand{\chem}[1]{\ensuremath{\mathrm{#1}}}

\begin{abstract}
We report results of the first muon-spin rotation measurements performed on the
low carrier density ferromagnet \chem{EuB_6}. The ferromagnetic state is
reached via two magnetic transitions at $T_{\rm m}=15.5$~K and $T_{\rm
c}=12.6$~K. Two distinct components are resolved in the muon data, one
oscillatory and one non-oscillatory, which arise from different types of
magnetic environment, and we have followed the temperature dependence of these
components in detail. These results provide evidence for magnetic phase
separation and can be interpreted in terms of the gradual coalescing of
magnetic polarons.
\end{abstract}


\maketitle

Europium hexaboride has attracted recent interest because it exhibits colossal
magnetoresistance (CMR) \cite{J_Appl_Phys_50_1911_1979} and it has been
suggested that its semiconductor-semimetal transition results from the overlap
of magnetic polarons \cite{Sullow:Metallization}. \chem{EuB_6} crystallises
into a simple cubic structure (space-group {\it Pm3m}) with divalent Eu ions
($^{8}S_{7\!/2}$) at the corners of the unit cell and B$_{6}$-octahedra at the
body-centred positions, and is a ferromagnet at low temperatures
\cite{J_Appl_Phys_50_1911_1979}.  Specific heat and magnetization measurements
reveal that this state is reached via two distinct transitions at $T_\mathrm{m}
= 15.5\,\mathrm{K}$ and $T_\mathrm{c} = 12.6\,\mathrm{K}$
\cite{Sullow:Metallization, Sullow:Struct_mag_order}. Neutron diffraction
measurements show that a small spontaneous magnetic moment begins to grow on
cooling below $T_\mathrm{m}$, but does not become significant until
$T_\mathrm{c}$ is reached, below which point the moment shows a more usual
mean-field like behaviour \cite{Neutrons}. The magnetic ordering is accompanied
by a sharp drop in the resistivity which is strongly field dependent
\cite{SSC_33_1055_1980} and gives rise to a large negative magnetoresistance
\cite{Sullow:Metallization}. This transition from a semiconductor at high
temperatures to a semimetal \cite{band_structure, SdH_and_dHvA, dHvA, Kondo} (or
possibly a self-doped compensated semiconductor \cite{Wigger:transport}) at low
temperatures is reminiscent of the metal-insulator transition seen in
manganites exhibiting CMR \cite{CMR_review}. Detailed measurements of
resistivity and magnetization \cite{Sullow:Metallization} show that this
transition is associated with $T_\mathrm{m}$. It is thought that magnetic
polarons could be responsible for this behaviour, with the upper magnetic
transition and drop in resistivity caused when the bound carriers overlap and
percolate, and the lower transition caused by a true transition to a bulk
ferromagnetic state \cite{Sullow:Metallization}. Further support for this explanation comes from
the observation of polaronic features, possibly associated with
itinerant holes\cite{maria}, below $\sim 30\,\mathrm{K}$ in
Raman-scattering spectra \cite{Raman}. The trapping of carriers to form bound
magnetic polarons provides an explanation for the upturn in resistivity
observed on cooling through $30\,\mathrm{K}$ \cite{Sullow:Metallization,maria} and the negative magnetoresistance \cite{Chatterjee:polarons, maria}.  However, a
direct observation of overlapping polarons in \chem{EuB_6} has so far been
elusive.

In this paper we present the results of $\mu$SR experiments on \chem{EuB_6}
which not only provide further evidence for the two distinct magnetic
transitions but are also able to resolve two components arising from muons
stopping in two different types of environment below $T_\mathrm{m}$. The first
component is an oscillating signal and can be associated with muons that stop
in a locally ferromagnetic environment. The second
component is a Gaussian signal and arises from the muons that come to rest in
a paramagnetic environment. This provides clear evidence for magnetic phase
separation below $T_{\mathrm{m}}$.

Our $\mu$SR experiments were carried out using the DOLLY instrument at the Paul
Scherrer Institute (PSI) in Switzerland and the DEVA beamline at the ISIS
pulsed muon facility in the UK.  In our $\mu$SR experiments, spin polarised
positive muons ($\mu^+$, mean lifetime $2.2\,\mu s$, momentum 28~MeV$/c$) were
implanted into polycrystalline EuB$_6$. The muons stop quickly (in
$<10^{-9}$~s), without significant loss of spin-polarisation. The time
evolution of the muon spin polarisation can be detected by counting emitted
decay positrons forward (f) and backward (b) of the initial muon spin direction
due to the asymmetric nature of the muon decay \cite{musr}.  In our experiments,
positrons are detected using scintillation counters placed in front of and
behind the sample.  We record the number of positrons detected by forward
($N_{\rm{f}}$) and backward ($N_{\rm{b}}$) counters as a function of time and
calculate the asymmetry function, $G_{z}(t)$, using

\begin{equation}
G_{z}(t)=\frac{N_{\rm{f}}(t)-\alpha_{\rm exp}
N_{\rm{b}}(t)}{N_{\rm{f}}(t)+\alpha_{\rm exp} N_{\rm{b}}(t)} ,
\label{asymmetry}
\end{equation}
where $\alpha_{\rm exp}$ is an experimental calibration constant and differs
from unity due to non-uniform detector efficiency. The quantity $G_{z}(t)$ is
then proportional to the average spin polarisation, $P_{z}(t)$, of muons
stopping within the sample. The muon spin precesses around a local magnetic
field, $B$ (with a frequency $\nu=(\gamma_{\mu}/2\pi) \vert B \vert$, where
$\gamma_{\mu}/2\pi= 135.5~\mathrm{MHz\,T}^{-1}$).

\par

\begin{figure}
\includegraphics[width=8.5cm]{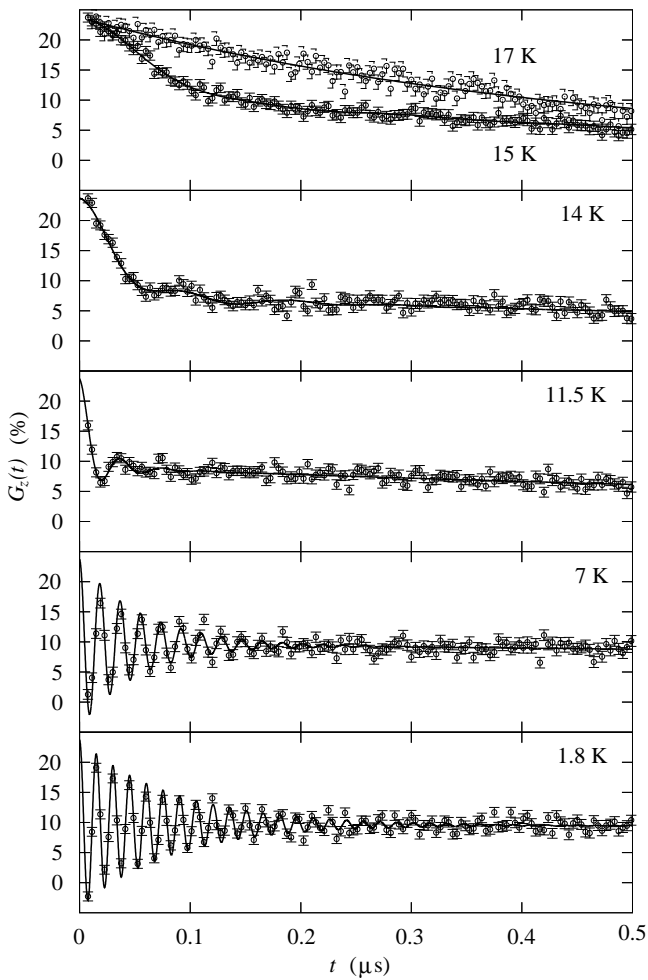}
\caption{Muon decay asymmetry plots for \chem{EuB_6} at different temperatures.
The solid lines are fits of the data to Eq.~\ref{fit_func}. The data for
$T=11.5, 14$ and $15\,\mathrm{K}$ are fitted with a non-zero Gaussian amplitude,
whilst the others are not. The data were taken at PSI. \label{EuB6_time}}
\end{figure}

Examples of asymmetry spectra measured at PSI are shown in
Fig.~\ref{EuB6_time}.  There are three distinct temperature regions.  In the
lowest temperature data ($T \lesssim 10\,\mathrm{K}$), there are clear
oscillations in the measured asymmetry, demonstrating that the sample does
indeed make a transition to a locally magnetically ordered state. This
oscillating signal is superposed on a slow exponential relaxation unobservable
in Fig.~\ref{EuB6_time}, but visible at longer times. In the second region
($10\,\mathrm{K} \lesssim T < T_{\mathrm{m}}$), the amplitude of the
oscillatory component decreases and a Gaussian component appears, with a decay
rate that decreases as the temperature is increased. In the third region ($T >
T_{\mathrm{m}}$), the amplitude associated with the Gaussian component
decreases until, when above $16\,\mathrm{K}$, only the slow exponential
relaxation remains. The need for the fast relaxing Gaussian term in the fits is
clearly motivated by the topmost panel of Fig.~\ref{EuB6_time}, where the
purely exponential relaxation seen at $17\,\mathrm{K}$ is compared with the
results for a temperature of $15\,\mathrm{K}$, which is just inside the
intermediate region.

\begin{figure}
\includegraphics[width=8.5cm]{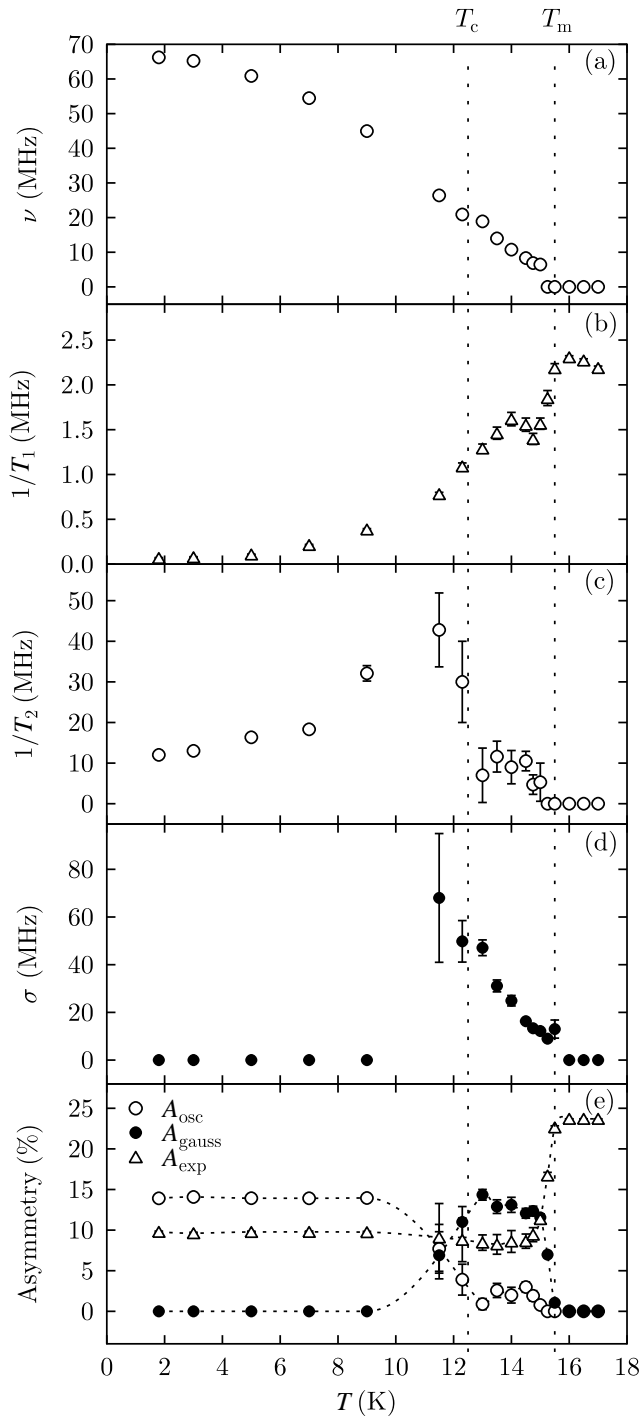}
\caption{Temperature dependence of the parameters determined from fits of
Eq.~\ref{fit_func} to \chem{EuB_6} asymmetry spectra.  The panels correspond to
(a) the oscillation frequency, $\nu$, (b) and (c) the relaxation rates
${1}/{T_1}$ and ${1}/{T_2}$, (d) the Gaussian fraction width parameter,
$\sigma$, and (e) the amplitudes for each of the three components used in the
fits.  The dashed lines in (e) are guides for the eye. The vertical dashed lines indicate the
positions of $T_\mathrm{c}$ and $T_\mathrm{m}$.  \label{EuB6_fitted}}
\end{figure}

In order to best follow these changes, the data over the whole studied
temperature range were fitted to the function
\begin{eqnarray}
G_{z}(t) &=& A_{\mathrm{exp}} \exp \left(-\frac{t}{T_{1}} \right) + A_{\mathrm{osc}}\exp \left(-\frac{t}{T_{2}} \right) \cos (2\pi \nu t) \nonumber \\
&+& A_{\mathrm{gauss}} \exp(-\sigma^2 t^2) + A_{\mathrm{bg}},
\label{fit_func}
\end{eqnarray}
where $A_{\mathrm{bg}}$ represents a time-independent background due to muons stopping
in the silver that surrounds the sample, $T_{1}$ and $T_{2}$ are the
longitudinal and transverse relaxation times, and $A_{\mathrm{exp}}$,
$A_{\mathrm{osc}}$ and $A_{\mathrm{gauss}}$ are the amplitudes of the
exponential, oscillating and Gaussian terms respectively. Plots of the fitted
parameters against temperature are shown in Fig.~\ref{EuB6_fitted}. Vertical
lines drawn at $T_\mathrm{m}$ and $T_\mathrm{c}$ approximately divide the plots
into the three temperature regions discussed above. In
Fig.~\ref{EuB6_fitted}(e) the asymmetry amplitudes of the three different terms
are shown, which provide the main motivation for this division. The
exponentially relaxing fraction is seen to drop sharply from the maximum value
at $T_{\mathrm{m}}$; this is expected in a polycrystalline sample when a
transition into an ordered state occurs. In the lowest temperature region the
remaining amplitude is accounted for by the oscillating fraction, so that here
the sample is uniformly magnetically ordered, but at
intermediate temperatures there is also a contribution from the fast relaxing
Gaussian fraction.

The fact that the asymmetry amplitude is shared between the two fast signals in
the intermediate region is evidence that the muons are stopping in two
different kinds of environment, that is, magnetic phase separation is occurring
\cite{phase_sep, char_spin_mod}. The amplitudes of the two signals are expected to be
proportional to the volume fraction of each phase. Phase separation is important
between $T_{\mathrm{m}}$ and $T_{\mathrm{c}}$, but we find it disappears below $T_{\mathrm{c}}$, in contrast to the predictions of Ref.~\onlinecite{char_spin_mod}.

A Gaussian relaxation can result from a field at the muon site which is static
but randomly distributed in magnitude. The parameter $\sigma$ is related to the
width of the field distribution as $\sigma^2 = {\gamma_{\mu}}^2 \langle B^2
\rangle / 2$. Fig.~\ref{EuB6_fitted}(d) shows that $\sigma$ increases as the
temperature is lowered in the second region, indicating that the field
distribution at the muon sites is becoming wider as the sample becomes more
ordered at lower temperatures. 

Fig.~\ref{EuB6_fitted}(a) shows that the oscillations develop below $T_{\rm
m}$, and their frequency $\nu$ increases fairly smoothly as the sample is
cooled through $T_{\rm c}$ rising to a maximum of $\sim$ 67~MHz at low
temperature (corresponding to a field at the muon site of $\sim 0.5$~T).  The
frequency $\nu$ is proportional to the magnetization and the data in
Fig.~\ref{EuB6_fitted}(a) match well with the temperature dependence of the Eu
magnetic moment measured with neutron scattering
\cite{Sullow:Struct_mag_order}.  Only one muon precession frequency is
observed, strongly suggesting that there is a single set of equivalent muon
sites in the structure. There are two candidate sites.  The first is at the
centre of a B$_6^{2-}$ octahedron, and the second is at the face-centres of the
unit cell (in the centre of the shortest B--B bond which is between atoms in
adjacent unit cells). Using the Eu moment measured previously
\cite{Sullow:Struct_mag_order} and assuming a low temperature magnetic
structure with the moments pointing along the [111] direction, the dipole field
($\boldsymbol{B}_{\mathrm{dip}}$) can be calculated at both these possible
sites. At the centre of a boron octahedron, the dipole field cancels by
symmetry.  The face-centre positions are all magnetically equivalent and yield
$\gamma_\mu\vert\boldsymbol{B}_{\rm dip}\vert/2\pi = 144\,\mathrm{MHz}$.
Additional contributions to the field at the muon site arise from the Lorentz
field ($\mu_{0}M_{\mathrm{sat}}/3 = 0.42\,\mathrm{T}$, corresponding to
$57\,\mathrm{MHz}$), the demagnetization field and the hyperfine contact field,
and preclude a definitive assignment of the site. 

The longitudinal relaxation rate, ${1}/{T_1}$, reflects the dynamics of the
fields being probed. For rapid fluctuations, $1/{T_{1}} \propto
\gamma^{2}_{\mu}\sum_{q}|\delta B(q)|^{2}\tau(q)$, where $|\delta B(q)|$ is the
amplitude of the fluctuating local field and $\tau(q)$ is the Eu-ion
correlation time at wavevector $q$.  In the paramagnetic phase the spin
fluctuations are so rapid that the measured relaxation rate is small. As the
sample is cooled and the critical region is approached, the correlation time
becomes longer and the relaxation rate rises and peaks close to
$T_{\rm m}$ (Fig.~\ref{EuB6_fitted}(b)). In contrast, ${1}/{T_2}$ (which is
proportional to the width of the ordered field distribution corresponding to
the oscillating fraction) rises dramatically on cooling through $T_{\rm c}$ and
subsequently falls on further cooling (Fig.~\ref{EuB6_fitted}(c)). This
emphasises that the ordered and fluctuating fractions in the sample are
distinct and follow different temperature dependences.
 
These observations fit in well with the polaron percolation picture for the
intermediate temperature range; muons stopping in ferromagnetic regions of
overlapping polarons give rise to the oscillating signal, while most muons stop
in the intervening paramagnetic regions. In these paramagnetic regions the
Eu-ion correlation time is short, so the muons are not depolarized by the local
fluctuating Eu moments, but by the distribution of fields that result from the
nearby ferromagnetic regions. As the sample becomes more ordered the
paramagnetic regions shrink, with a corresponding effect on the field
distribution.

\begin{figure}
\includegraphics[width=8.5cm]{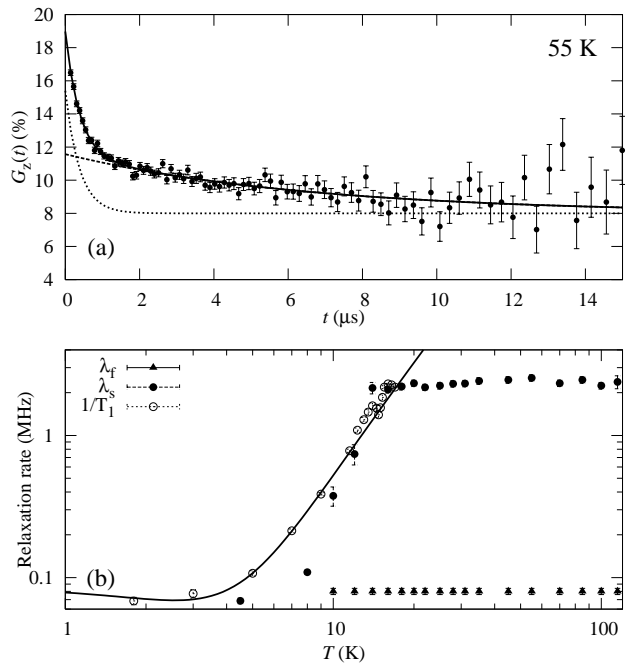}
\caption{Panel (a) shows a typical asymmetry spectrum measured at ISIS. The
solid line is a fit to Eq.~\ref{eq:ISIS}, and the dotted lines show the form of
the two exponential terms used. Panel (b) shows the two relaxation rates
measured at ISIS, $\lambda_{\mathrm{f,s}}$ (filled symbols), and the
longitudinal relaxation rate measured at PSI, $1/T_{1}$ (open symbols).  The
solid line is a fit of a $T^{2}\ln(T/\Delta)$ function to the PSI rate for
temperatures below $14\,\mathrm{K}$.  \label{EuB6_ISIS}}
\end{figure}

In order to study the slower relaxation in the high temperature region in more
detail, data were collected at the ISIS facility. An example is shown in
Fig.~\ref{EuB6_ISIS}(a) and clearly shows the presence of both a slow and a
fast relaxation rate.  Therefore the data were fitted to the function
\cite{ref:Heffner_two_exp}
\begin{equation}
G_z(t) = A_{\rm f} \exp(-\lambda_{\rm f}t) + A_{\rm s}\exp(-\lambda_{\rm
 s}t) 
+ A_{\mathrm{bg}},
\label{eq:ISIS}
\end{equation}
where $A_{\rm f,s}$ and $\lambda_{\rm f,s}$ are the amplitudes and relaxation
rates of the fast and slow exponential components (note that $\lambda_{\rm f}$
represents relaxation due to {\it fast} dynamics and hence {\it slow} spin
relaxation).  The non-relaxing background fraction $A_{\mathrm{bg}}$ was held
fixed. $A_{\rm f}$ and $A_{\rm s}$ were found to be approximately constant with
temperature, with $A_{\rm f}$ about half the value of $A_{\rm s}$, and were
fixed in the fits. The relaxation rates themselves appear temperature
independent above the transition, and it was possible to keep $\lambda_{\rm f}$
fixed at $0.08\,\mathrm{MHz}$ (see Fig.~\ref{EuB6_ISIS}). On cooling
towards $T_{\rm c}$, $\lambda_{\rm s}$ begins to decrease, until only one
relaxation rate can be resolved below $9\,\mathrm{K}$. This is possibly because
the two rates become very similar in magnitude, and the presence of
$\lambda_{\rm f}$ may mask any further drop of $\lambda_{\rm s}$.

The relaxation rate $\lambda_{\rm s}$ matches well with the temperature
dependence of $1/T_{1}$, as shown in Fig.~\ref{EuB6_ISIS}(b), and the two can
be identified with each other.  At low temperatures, a contribution to $1/T_1$
can be fitted by $1/T_1 \propto T^{2}\ln\left( T/\Delta \right)$ (see
Fig.~\ref{EuB6_ISIS}(b)), appropriate for scattering by two-magnon processes in
ferromagnets \cite{lowT_ferro_relax}.  However, the component due to fast
fluctuations produces too slow a relaxation to be convincingly included in the
fits to the PSI data. Nevertheless, the observation of two relaxation rates
above $T_{\rm m}$ allows us to infer the presence of spatial inhomogeneity (a
very similar effect has been found in \chem{La_{0.67}Ca_{0.33}MnO_3} with
$\mu$SR \cite{ref:Heffner_two_exp}), which persists even at $115\,\mathrm{K}$.
Above $T_{\rm m}$, the temperature dependence of $\lambda_{\rm f}$ and
$\lambda_{\rm s}$ is weak and featureless.  Any polaron dynamics are presumably
too fast to be followed by the muon.  The dramatic changes observed below
$T_{\rm m}$ can therefore be attributed to a large change in the time-scale of
polaron dynamics, such as might be expected when polarons overlap. Although the
polaron volume fraction changes little at a percolation transition, their
arrangement could be changed so that the average size of a single polaron is
larger and its dynamics are much slower.

In conclusion, $\mu$SR measurements have allowed us to follow the very unusual
development of ferromagnetism in \chem{EuB_6} through the transitions at
$T_\mathrm{m}$ and $T_{\mathrm{c}}$ from a local viewpoint.  These results
reveal two distinct and spatially separate regions of the material associated
with different magnetic behaviour. Such magnetic phase separation is qualitatively consistent with a picture based on coalescing polarons.

Part of this work was performed at the Swiss Muon Source, Paul Scherrer
Institute, Villigen, Switzerland and the ISIS pulsed muon source, RAL, UK. We
are grateful to Robert Scheuermann and Steve Cottrell for experimental
assistance and Maria Calder\'on and Amalia Coldea for useful discussions.  This
work was funded by the EPSRC (UK) and the NSF under grant DMR-0203214

\bibliography{EuB6_Muon}

\begin{thebibliography}{19}
\expandafter\ifx\csname natexlab\endcsname\relax\def\natexlab#1{#1}\fi
\expandafter\ifx\csname bibnamefont\endcsname\relax
  \def\bibnamefont#1{#1}\fi
\expandafter\ifx\csname bibfnamefont\endcsname\relax
  \def\bibfnamefont#1{#1}\fi
\expandafter\ifx\csname citenamefont\endcsname\relax
  \def\citenamefont#1{#1}\fi
\expandafter\ifx\csname url\endcsname\relax
  \def\url#1{\texttt{#1}}\fi
\expandafter\ifx\csname urlprefix\endcsname\relax\def\urlprefix{URL }\fi
\providecommand{\bibinfo}[2]{#2}
\providecommand{\eprint}[2][]{\url{#2}}

\bibitem[{\citenamefont{Fisk et~al.}(1979)\citenamefont{Fisk, Johnston, Cornut,
  von Molnar, Oseroff, and Calvo}}]{J_Appl_Phys_50_1911_1979}
\bibinfo{author}{\bibfnamefont{Z.}~\bibnamefont{Fisk}},
  \bibinfo{author}{\bibfnamefont{D.~C.} \bibnamefont{Johnston}},
  \bibinfo{author}{\bibfnamefont{B.}~\bibnamefont{Cornut}},
  \bibinfo{author}{\bibfnamefont{S.}~\bibnamefont{von Molnar}},
  \bibinfo{author}{\bibfnamefont{S.}~\bibnamefont{Oseroff}}, \bibnamefont{and}
  \bibinfo{author}{\bibfnamefont{R.}~\bibnamefont{Calvo}}, \bibinfo{journal}{J.
  Appl. Phys.} \textbf{\bibinfo{volume}{50}}, \bibinfo{pages}{1911}
  (\bibinfo{year}{1979}).

\bibitem[{\citenamefont{S\"{u}llow et~al.}(2000)\citenamefont{S\"{u}llow,
  Prasad, Aronson, Bogdanovich, Sarrao, and Fisk}}]{Sullow:Metallization}
\bibinfo{author}{\bibfnamefont{S.}~\bibnamefont{S\"{u}llow}},
  \bibinfo{author}{\bibfnamefont{I.}~\bibnamefont{Prasad}},
  \bibinfo{author}{\bibfnamefont{M.~C.} \bibnamefont{Aronson}},
  \bibinfo{author}{\bibfnamefont{S.}~\bibnamefont{Bogdanovich}},
  \bibinfo{author}{\bibfnamefont{J.~L.} \bibnamefont{Sarrao}},
  \bibnamefont{and} \bibinfo{author}{\bibfnamefont{Z.}~\bibnamefont{Fisk}},
  \bibinfo{journal}{Phys. Rev. B} \textbf{\bibinfo{volume}{62}},
  \bibinfo{pages}{11626} (\bibinfo{year}{2000}).

\bibitem[{\citenamefont{S{\"u}llow et~al.}(1998)\citenamefont{S{\"u}llow,
  Prasad, Aronson, Sarrao, Fisk, Hristova, Lacerda, Hundley, Vigliante, and
  Gibbs}}]{Sullow:Struct_mag_order}
\bibinfo{author}{\bibfnamefont{S.}~\bibnamefont{S{\"u}llow}},
  \bibinfo{author}{\bibfnamefont{I.}~\bibnamefont{Prasad}},
  \bibinfo{author}{\bibfnamefont{M.~C.} \bibnamefont{Aronson}},
  \bibinfo{author}{\bibfnamefont{J.~L.} \bibnamefont{Sarrao}},
  \bibinfo{author}{\bibfnamefont{Z.}~\bibnamefont{Fisk}},
  \bibinfo{author}{\bibfnamefont{D.}~\bibnamefont{Hristova}},
  \bibinfo{author}{\bibfnamefont{A.~H.} \bibnamefont{Lacerda}},
  \bibinfo{author}{\bibfnamefont{M.~F.} \bibnamefont{Hundley}},
  \bibinfo{author}{\bibfnamefont{A.}~\bibnamefont{Vigliante}},
  \bibnamefont{and} \bibinfo{author}{\bibfnamefont{D.}~\bibnamefont{Gibbs}},
  \bibinfo{journal}{Phys. Rev. B} \textbf{\bibinfo{volume}{57}},
  \bibinfo{pages}{5860} (\bibinfo{year}{1998}).

\bibitem[{\citenamefont{Henggeler et~al.}(1998)\citenamefont{Henggeler, Ott,
  Young, and Fisk}}]{Neutrons}
\bibinfo{author}{\bibfnamefont{W.}~\bibnamefont{Henggeler}},
  \bibinfo{author}{\bibfnamefont{H.-R.} \bibnamefont{Ott}},
  \bibinfo{author}{\bibfnamefont{D.~P.} \bibnamefont{Young}}, \bibnamefont{and}
  \bibinfo{author}{\bibfnamefont{Z.}~\bibnamefont{Fisk}},
  \bibinfo{journal}{Solid State Commun.} \textbf{\bibinfo{volume}{108}},
  \bibinfo{pages}{929} (\bibinfo{year}{1998}).

\bibitem[{\citenamefont{Guy et~al.}(1980)\citenamefont{Guy, v.~Molnar,
  Etourneau, and Fisk}}]{SSC_33_1055_1980}
\bibinfo{author}{\bibfnamefont{C.~N.} \bibnamefont{Guy}},
  \bibinfo{author}{\bibfnamefont{S.}~\bibnamefont{v.~Molnar}},
  \bibinfo{author}{\bibfnamefont{J.}~\bibnamefont{Etourneau}},
  \bibnamefont{and} \bibinfo{author}{\bibfnamefont{Z.}~\bibnamefont{Fisk}},
  \bibinfo{journal}{Solid State Commun.} \textbf{\bibinfo{volume}{33}},
  \bibinfo{pages}{1055} (\bibinfo{year}{1980}).

\bibitem[{\citenamefont{Massidda et~al.}(1996)\citenamefont{Massidda,
  Continenza, de~Pascale, and Monnier}}]{band_structure}
\bibinfo{author}{\bibfnamefont{S.}~\bibnamefont{Massidda}},
  \bibinfo{author}{\bibfnamefont{A.}~\bibnamefont{Continenza}},
  \bibinfo{author}{\bibfnamefont{T.~M.} \bibnamefont{de~Pascale}},
  \bibnamefont{and} \bibinfo{author}{\bibfnamefont{R.}~\bibnamefont{Monnier}},
  \bibinfo{journal}{Z. Phys. B: Condensed Matter}
  \textbf{\bibinfo{volume}{102}}, \bibinfo{pages}{83} (\bibinfo{year}{1996}).

\bibitem[{\citenamefont{Aronson et~al.}(1999)\citenamefont{Aronson, Sarrao,
  Fisk, Whitton, and Brandt}}]{SdH_and_dHvA}
\bibinfo{author}{\bibfnamefont{M.~C.} \bibnamefont{Aronson}},
  \bibinfo{author}{\bibfnamefont{J.~L.} \bibnamefont{Sarrao}},
  \bibinfo{author}{\bibfnamefont{Z.}~\bibnamefont{Fisk}},
  \bibinfo{author}{\bibfnamefont{M.}~\bibnamefont{Whitton}}, \bibnamefont{and}
  \bibinfo{author}{\bibfnamefont{B.~L.} \bibnamefont{Brandt}},
  \bibinfo{journal}{Phys. Rev. B} \textbf{\bibinfo{volume}{59}},
  \bibinfo{pages}{4720} (\bibinfo{year}{1999}).

\bibitem[{\citenamefont{Goodrich et~al.}(1998)\citenamefont{Goodrich, Harrison,
  Vuillemin, Teklu, Hall, Fisk, Young, and Sarrao}}]{dHvA}
\bibinfo{author}{\bibfnamefont{R.~G.} \bibnamefont{Goodrich}},
  \bibinfo{author}{\bibfnamefont{N.}~\bibnamefont{Harrison}},
  \bibinfo{author}{\bibfnamefont{J.~J.} \bibnamefont{Vuillemin}},
  \bibinfo{author}{\bibfnamefont{A.}~\bibnamefont{Teklu}},
  \bibinfo{author}{\bibfnamefont{D.~W.} \bibnamefont{Hall}},
  \bibinfo{author}{\bibfnamefont{Z.}~\bibnamefont{Fisk}},
  \bibinfo{author}{\bibfnamefont{D.}~\bibnamefont{Young}}, \bibnamefont{and}
  \bibinfo{author}{\bibfnamefont{J.}~\bibnamefont{Sarrao}},
  \bibinfo{journal}{Phys. Rev. B} \textbf{\bibinfo{volume}{58}},
  \bibinfo{pages}{14896} (\bibinfo{year}{1998}).

\bibitem[{\citenamefont{Kun\v{e} and Pickett}(2004)}]{Kondo}
\bibinfo{author}{\bibfnamefont{J.}~\bibnamefont{Kun\v{e}}} \bibnamefont{and}
  \bibinfo{author}{\bibfnamefont{W.~E.} \bibnamefont{Pickett}},
  \bibinfo{journal}{Phys. Rev. B} \textbf{\bibinfo{volume}{69}},
  \bibinfo{pages}{165111} (\bibinfo{year}{2004}).

\bibitem[{\citenamefont{Wigger et~al.}(2004)\citenamefont{Wigger, Monnier, Ott,
  Young, and Fisk}}]{Wigger:transport}
\bibinfo{author}{\bibfnamefont{G.~A.} \bibnamefont{Wigger}},
  \bibinfo{author}{\bibfnamefont{R.}~\bibnamefont{Monnier}},
  \bibinfo{author}{\bibfnamefont{H.~R.} \bibnamefont{Ott}},
  \bibinfo{author}{\bibfnamefont{D.~P.} \bibnamefont{Young}}, \bibnamefont{and}
  \bibinfo{author}{\bibfnamefont{Z.}~\bibnamefont{Fisk}},
  \bibinfo{journal}{Phys. Rev. B} \textbf{\bibinfo{volume}{69}},
  \bibinfo{pages}{125118} (\bibinfo{year}{2004}).

\bibitem[{\citenamefont{Imada et~al.}(1998)\citenamefont{Imada, Fujimori, and
  Tokura}}]{CMR_review}
\bibinfo{author}{\bibfnamefont{M.}~\bibnamefont{Imada}},
  \bibinfo{author}{\bibfnamefont{A.}~\bibnamefont{Fujimori}}, \bibnamefont{and}
  \bibinfo{author}{\bibfnamefont{Y.}~\bibnamefont{Tokura}},
  \bibinfo{journal}{Rev. Mod. Phys.} \textbf{\bibinfo{volume}{70}},
  \bibinfo{pages}{1039} (\bibinfo{year}{1998}).

\bibitem[{\citenamefont{Calder\'on et~al.}(2003)\citenamefont{Calder\'on,
  Wegener, and Littlewood}}]{maria}
\bibinfo{author}{\bibfnamefont{M.~J.} \bibnamefont{Calder\'on}},
  \bibinfo{author}{\bibfnamefont{L.~G.~L.} \bibnamefont{Wegener}},
  \bibnamefont{and} \bibinfo{author}{\bibfnamefont{P.~B.}
  \bibnamefont{Littlewood}}, \bibinfo{journal}{cond-mat/0312437}
  (\bibinfo{year}{2003}).

\bibitem[{\citenamefont{Nyhus et~al.}(1997)\citenamefont{Nyhus, Yoon, Kauffman,
  Cooper, Fisk, and Sarrao}}]{Raman}
\bibinfo{author}{\bibfnamefont{P.}~\bibnamefont{Nyhus}},
  \bibinfo{author}{\bibfnamefont{S.}~\bibnamefont{Yoon}},
  \bibinfo{author}{\bibfnamefont{M.}~\bibnamefont{Kauffman}},
  \bibinfo{author}{\bibfnamefont{S.~L.} \bibnamefont{Cooper}},
  \bibinfo{author}{\bibfnamefont{Z.}~\bibnamefont{Fisk}}, \bibnamefont{and}
  \bibinfo{author}{\bibfnamefont{J.}~\bibnamefont{Sarrao}},
  \bibinfo{journal}{Phys. Rev. B} \textbf{\bibinfo{volume}{56}},
  \bibinfo{pages}{2717} (\bibinfo{year}{1997}).

\bibitem[{\citenamefont{Chatterjee et~al.}(2004)\citenamefont{Chatterjee, Yu,
  and Min}}]{Chatterjee:polarons}
\bibinfo{author}{\bibfnamefont{J.}~\bibnamefont{Chatterjee}},
  \bibinfo{author}{\bibfnamefont{U.}~\bibnamefont{Yu}}, \bibnamefont{and}
  \bibinfo{author}{\bibfnamefont{B.~I.} \bibnamefont{Min}},
  \bibinfo{journal}{Phys. Rev. B} \textbf{\bibinfo{volume}{69}},
  \bibinfo{pages}{134423} (\bibinfo{year}{2004}).

\bibitem[{\citenamefont{Blundell}(1999)}]{musr}
\bibinfo{author}{\bibfnamefont{S.~J.} \bibnamefont{Blundell}},
  \bibinfo{journal}{Comtemp. Phys.} \textbf{\bibinfo{volume}{40}},
  \bibinfo{pages}{175} (\bibinfo{year}{1999}).

\bibitem[{\citenamefont{Kim et~al.}(2002)\citenamefont{Kim, Uehara, Kiryukhin,
  and Cheong}}]{phase_sep}
\bibinfo{author}{\bibfnamefont{K.~H.} \bibnamefont{Kim}},
  \bibinfo{author}{\bibfnamefont{M.}~\bibnamefont{Uehara}},
  \bibinfo{author}{\bibfnamefont{V.}~\bibnamefont{Kiryukhin}},
  \bibnamefont{and} \bibinfo{author}{\bibfnamefont{S.-W.}
  \bibnamefont{Cheong}}, \bibinfo{journal}{cond-mat/0212113}
  (\bibinfo{year}{2002}).

\bibitem[{\citenamefont{Korenblit}(2001)}]{char_spin_mod}
\bibinfo{author}{\bibfnamefont{I.~Y.} \bibnamefont{Korenblit}},
  \bibinfo{journal}{Phys. Rev. B} \textbf{\bibinfo{volume}{64}},
  \bibinfo{pages}{100405} (\bibinfo{year}{2001}).

\bibitem[{\citenamefont{Heffner et~al.}(2000)\citenamefont{Heffner, Sonier,
  MacLaughlin, Nieuwenhuys, Ehlers, Mezei, Cheong, Gardner, and
  R\"{o}der}}]{ref:Heffner_two_exp}
\bibinfo{author}{\bibfnamefont{R.~H.} \bibnamefont{Heffner}},
  \bibinfo{author}{\bibfnamefont{J.~E.} \bibnamefont{Sonier}},
  \bibinfo{author}{\bibfnamefont{D.~E.} \bibnamefont{MacLaughlin}},
  \bibinfo{author}{\bibfnamefont{G.~J.} \bibnamefont{Nieuwenhuys}},
  \bibinfo{author}{\bibfnamefont{G.}~\bibnamefont{Ehlers}},
  \bibinfo{author}{\bibfnamefont{F.}~\bibnamefont{Mezei}},
  \bibinfo{author}{\bibfnamefont{S.-W.} \bibnamefont{Cheong}},
  \bibinfo{author}{\bibfnamefont{J.~S.} \bibnamefont{Gardner}},
  \bibnamefont{and}
  \bibinfo{author}{\bibfnamefont{H.}~\bibnamefont{R\"{o}der}},
  \bibinfo{journal}{Phys. Rev. Lett.} \textbf{\bibinfo{volume}{85}},
  \bibinfo{pages}{3285} (\bibinfo{year}{2000}).

\bibitem[{\citenamefont{Yaouanc and de~R\'{e}otier}(1991)}]{lowT_ferro_relax}
\bibinfo{author}{\bibfnamefont{A.}~\bibnamefont{Yaouanc}} \bibnamefont{and}
  \bibinfo{author}{\bibfnamefont{P.~D.} \bibnamefont{de~R\'{e}otier}},
  \bibinfo{journal}{J. Phys.: Condens. Matter} \textbf{\bibinfo{volume}{3}},
  \bibinfo{pages}{6195} (\bibinfo{year}{1991}).

\end{thebibliography}

\end{document}